\documentclass[twocolumn,preprintnumbers,amsmath,amssymb,prb,longbibliography]{revtex4-2}
\usepackage{txfonts}
\usepackage{subfigure}
\usepackage{dcolumn}
\usepackage{bm}
\usepackage{braket}
\usepackage{upgreek}
\usepackage{extarrows}
\usepackage{float}
\usepackage{array}
\usepackage{booktabs} 
\usepackage{multirow}
\usepackage{graphicx}
\usepackage{bm}
\usepackage{subfigure,multirow}
\usepackage[T1]{fontenc}
\usepackage[mathscr]{euscript}
\usepackage{amsmath}
\graphicspath{ {./figures/} }
\usepackage{url}
\usepackage{graphicx}
\usepackage[colorlinks,linkcolor=blue,anchorcolor=blue,citecolor=blue]{hyperref}
\hypersetup{colorlinks=true,linkcolor=blue,citecolor=blue,urlcolor=blue}
\usepackage{color}

\begin{document}

\title{Necessity of orthogonal basis vectors for the two-anyon problem in one-dimensional lattice}

\author{Cuicui Zheng}
\affiliation{Key Laboratory of Quantum States of Matter and Optical Field Manipulation of Zhejiang Province, Department of Physics, Zhejiang Sci-Tech University, 310018 Hangzhou, China}
\author{Jiahui Xie}
\affiliation{Key Laboratory of Quantum States of Matter and Optical Field Manipulation of Zhejiang Province, Department of Physics, Zhejiang Sci-Tech University, 310018 Hangzhou, China}
\author{Ming Zhang}
\affiliation{Key Laboratory of Quantum States of Matter and Optical Field Manipulation of Zhejiang Province, Department of Physics, Zhejiang Sci-Tech University, 310018 Hangzhou, China}
\author{Yajiang Chen}
\email{yjchen@zstu.edu.cn}
\affiliation{Key Laboratory of Quantum States of Matter and Optical Field Manipulation of Zhejiang Province, Department of Physics, Zhejiang Sci-Tech University, 310018 Hangzhou, China}
\author{Yunbo Zhang}
\email{ybzhang@zstu.edu.cn}
\affiliation{Key Laboratory of Quantum States of Matter and Optical Field Manipulation of Zhejiang Province, Department of Physics, Zhejiang Sci-Tech University, 310018 Hangzhou, China}
\date{\today}
\begin{abstract}
Few-body physics for anyons has been intensively studied within the anyon-Hubbard model, including the quantum walk and Bloch oscillations of two-anyon states. However, the known theoretical proposal and experimental simulations of two-anyon states in one-dimensional lattice have been carried out by expanding the wavefunction in terms of non-orthogonal basis vectors, which introduces extra non-physical degrees of freedom. In the present work, we deduce the finite difference equations for the two-anyon state in the one-dimensional lattice by solving the Schr\"odinger equation with orthogonal basis vectors. Such an orthogonal scheme gives all the orthogonal physical eigenstates for the time-independent two-anyon Schr\"odinger equation, while the conventional (non-orthogonal) method produces a lot of non-physical redundant eigen-solutions whose components violate the anyonic relations. The dynamical property of the two-anyon states in a sufficiently large lattice has been investigated and compared in both the orthogonal and conventional schemes, which proves to depend crucially on the initial states. When the initial states with two anyons on the same site or (next-)neighboring sites are suitably chosen to be in accordance with the anyonic coefficient relation, we observe exactly the same dynamical behavior in the two schemes, including the revival probability, the probability density function, and the two-body correlation, otherwise, the conventional scheme will produce erroneous results which not any more describe anyons. The period of the Bloch oscillation in the pseudo-fermionic limit is found to be twice that in the bosonic limit, while the oscillations disappear for statistical parameters in between. Our findings are vital for quantum simulations of few-body physics with anyons in the lattice. 
\end{abstract}
\maketitle

\section{Introduction}
Over the years, exotic anyonic systems have continuously stimulated the research interests of theoretical and experimental physicists. As a kind of identical particle in low dimensions, anyons obey the fractional statistics, i.e., the wave function acquires a phase factor when exchanging two anyons~\cite{Laidlaw1971,Leinaas1977a,Wilczek1982,Lerda1992}. They often arise as quasi-particle excitations in two-dimensional condensed-matter systems, e.g., the fractional quantum Hall state~\cite{canright1990a,Laughlin1983,Halperin1984,Camino2005,Kim2005,stern2008,bartolomei2020a} and the topologically nontrivial chiral spin liquid~\cite{yao2007a,bauer2014}, or as virtual particles simulated in one dimensional systems~\cite{Ha1994,Murthy1994,Ha1995,Batchelor2006,Batchelor2007,Hao2008,Hao2009,Bromberg2009,greschner2015,Alicea2016,Zhang2017,rossini2019} and even in arbitrary dimensions~\cite{Haldane1991}. Especially, non-Abelian anyons have been constructed on a trapped-ion processor~\cite{iqbal2024}, which can be a promising building block of a fault-tolerant topological quantum computer~\cite{kitaev2003,Nayak2008,lahtinen2017}.

Due to the rapid advancement of controllable anyonic platforms~\cite{keilmann2011b,georgescu2014,nakamura2020,altman2021}, few-body physics for Abelian anyons has been intensively studied within the anyon-Hubbard model, including the quantum walk~\cite{Sansoni2012, campagnano2012, Wang2014, lau2022} and Bloch oscillations of two-anyon states~\cite{Longhi2012,corrielli2013,Zhang2022}. The quantum walk behavior of identical particles reflects the effects of the exchange statistics~\cite{hong1987} and can be a promising candidate to realize quantum computing~\cite{childs2009,childs2013}. The two-anyon quantum walk exhibits a statistics-dependent bunching/antibunching phenomenon in the two-body correlation functions~\cite{Sansoni2012} and spatially asymmetric transport~\cite{Wang2014,Kwan2023}. Even a statistical boundary, which is transparent to single anyons and exchanging multiple particles arriving together, has been introduced to engineer the two-anyon quantum walk~\cite{lau2022}. Under a linear external potential, the energy spectrum of ``bound'' two-anyon states in a sufficiently large lattice can have a uniform energy interval between neighboring states for both non-interacting bosons and pseudo-fermions~\cite{keilmann2011b,Wang2014}, and, the energy interval for pseudo-fermions is half of the case for bosons, which leads to the same relation between the frequencies of their Bloch oscillations (BOs)~\cite{Longhi2012}. Such anyonic BOs have been observed in the simulators using a photonic lattice~\cite{corrielli2013} and electric circuits~\cite{Zhang2022}.  

The theoretical studies on the two-anyon quantum walk in Refs.~\cite{Sansoni2012, Wang2014,lau2022} have ignored the energy spectrum of the system and only focused on the dynamical properties by solving the time-dependent Schr\"odinger equation based on the fractional Jordan-Wigner transformation~\cite{fradkin1989,keilmann2011b}, which maps anyons to interacting bosons. The information on the energy spectrum of few-body anyonic systems can be crucial in many circumstances, e.g., the anyonic Bloch oscillations~\cite{Longhi2012}, where another routine to obtain the dynamic behavior of few-body anyonic systems has employed the commutation relations of the anyonic creation and annihilation operators, and directly solved the time-dependent two-anyonic Schr\"odinger equation. However, the wave functions of the two-anyon states in Ref.~\cite{Longhi2012} have been expanded by a set of non-orthogonal basis vectors in the Fock space, which introduces extra non-physical degrees of freedom into the systems, e.g., non-physical eigenstates, and may in some cases result in some erroneous dynamical features. More seriously, this routine has been followed in the important experimental simulations~\cite{corrielli2013, Zhang2022}. Therefore, it is necessary and urgent to provide a rectification for the anyonic few-body physics in the anyon-Hubbard model.

In this work, we obtain the finite difference equations for the two-anyon state in the one-dimensional lattice by solving the Schr\"odinger equation with orthogonal basis vectors, which is vital for various anyonic quantum simulations. Our orthogonal scheme gives all the orthogonal physical eigenstates for the time-independent two-anyon Schr\"odinger equation, while the conventional (non-orthogonal) method produces a lot of extra non-physical eigen-solutions. By means of these orthogonal basis vectors we examine the dynamical behavior of the two-anyon state. Both schemes show that a two-anyon state in a sufficiently large lattice exhibits the anyonic BOs in the time-dependent revival probability and probability density function when the initial state is physical. However, for the "non-anyonic" initial states, the conventional scheme gives an incorrect prediction of the BO at the original period, leading to a probability density distribution violating the conservation law in the fermionic limit. 



The paper is organized as follows: In Sec. II we give wave function descriptions and the stationary eigenquations with respect to the coefficients in both conventional and orthogonal spaces for two anyons in the 1D lattice. We take a small-sized lattice as an example of the analytical solutions for the energy spectrum and then numerical results for a sufficiently large stationary system are illustrated in Sec. III. The dynamics of the two correlated anyons for a variety of typical initial states are discussed in Sec. IV for various statistical parameters. Concluding remarks are given in Sec. V.

\section{ Model }\label{II}

We consider a pair of non-interacting anyons hopping on a one-dimensional (1D) lattice under a uniform external force $F$. The corresponding Hamiltonian is given as~\cite{Longhi2012,Zhang2022}
\begin{equation}
    \label{Hamiltonian}
	H=-J \sum_{l=1}^{L}(a_{l}^{\dagger}a_{l+1}+a_{l+1}^{\dagger}a_{l})+F\sum_{l=1}^{L}l\,n_{l},
\end{equation}    
where the operator $a_{l}$ ($a^\dagger_{l}$) annihilates (creates) one anyon at site $l$, $n_{l}=a_{l}^{\dagger}a_{l}$ denotes the particle number operator, $L$ is the number of lattice sites, and $J$ is the tunneling amplitude between the nearest-neighbor sites. The anyonic operators satisfy the following commutation relations~\cite{Longhi2012}
\begin{eqnarray}
\label{CommutationRelation}
	a_{l}a_{k}^{\dagger}&=&e^{-i\chi \pi \epsilon(l-k)}a_{k}^{\dagger}a_{l}+\delta_{lk}, \notag\\
	a_{l}a_{k}&=&e^{i\chi \pi \epsilon(l-k)}a_{k}a_{l}, \\ 
	a^\dagger_{l}a^\dagger_{k}&=&e^{i\chi \pi \epsilon(l-k)}a^\dagger_{k}a^\dagger_{l}, \notag
\end{eqnarray}
where $\chi \in  [0,1] $ is the anyonic statistical parameter, and the sign function $\epsilon(x)$ gives $1,0$ or $-1$ depending on whether $x$ is positive, zero, or negative. Two anyons with $\chi=0$ behave as conventional bosons, while those with $\chi=1$ are pseudo-fermions which act like fermions off-site and bosons on-site{\color{red}}.

\begin{figure}[t]
	\centering
	\includegraphics[width=1\linewidth]{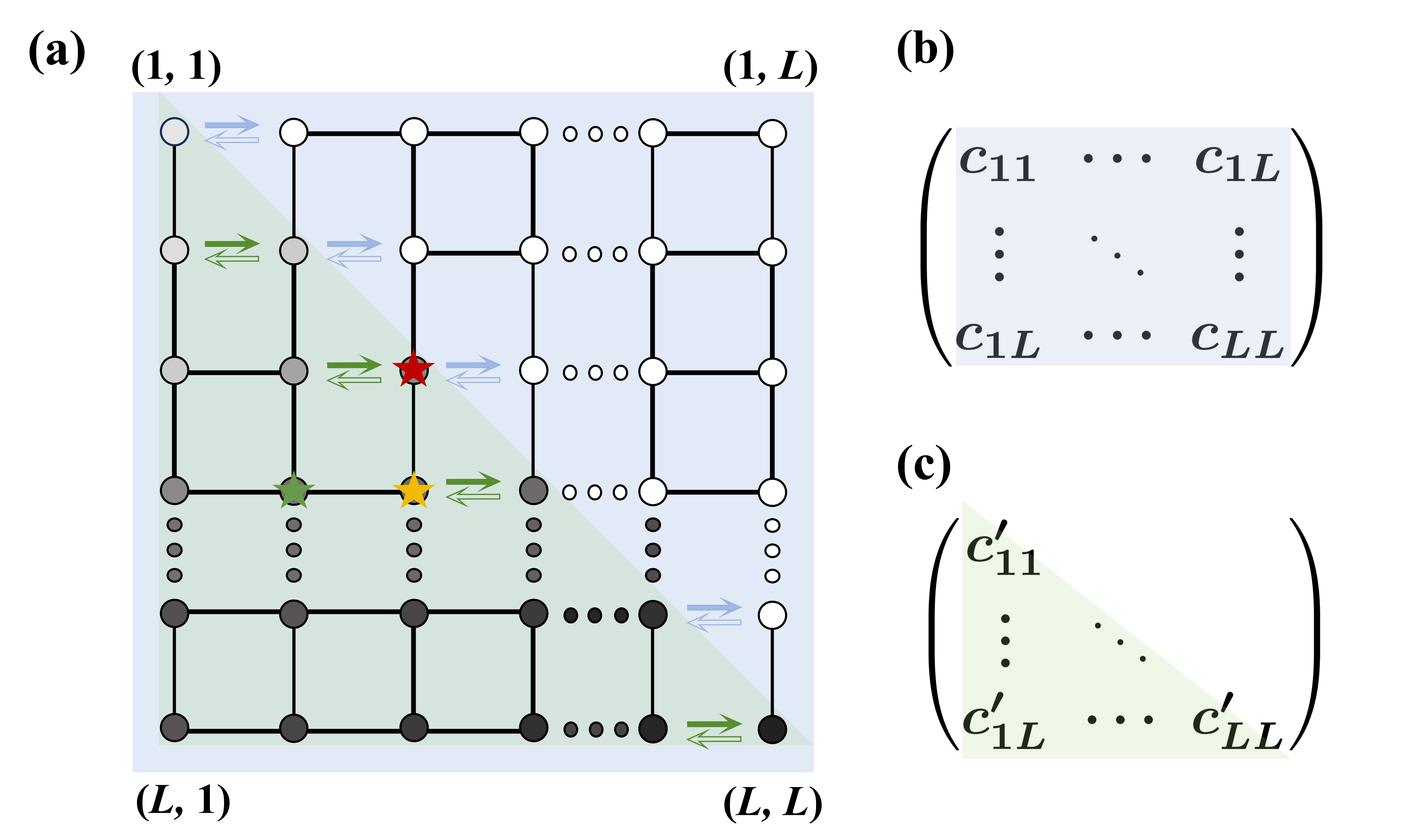}
	\caption{(Color online) (a) The schematic mapping of the 1D two-anyon system into a single particle motion on a 2D lattice in the presence of an external force, according to Eqs.~(\ref{coefficients}) and (\ref{coefficients22}). The entire square lattice constitutes the conventional Hilbert space with each site being a probability amplitude $c_{lk}$, while the folded triangular area in light green represents the orthogonal Hilbert space with $l\ge k$. The gradient gray of the sites shows the on-site potential energy given by the external force, and the left/right arrows correspond to jumps with complex phases $e^{\pm i\chi\pi}$. Panels (b) and (c) show the expansion coefficients in Eqs.~(\ref{eigenstates}) and (\ref{a3}) for the conventional and orthogonal Hilbert spaces with dimensions $L^2$ and $L(L+1)/2$, respectively. The colored stars indicate three initial states that we choose for the study of dynamical properties.
 } \label{fig1}
\end{figure}

The analysis of the eigenenergy spectrum and the BOs dynamical motion of two correlated anyons greatly depends on the choice of the basis vectors of the Hilbert space. Since $[N, H] = 0$ with $N$ being the total particle number $N=\sum_l n_l$, the system evolves in the two-particle Hilbert space. One conventional choice, employed in Refs.~\cite{Longhi2012,Zhang2022}, is the Fock states with one anyon at site $l$ and the other one at site $k$ of the lattice, i.e. $a_{l}^{\dagger}a_{k}^{\dagger}\left | 0  \right \rangle $ with $\left | 0  \right \rangle$ the vacuum state and $l,k=1,2,\cdots, L$. The eigenstates of the Hamiltonian Eq.~(\ref{Hamiltonian}) are expanded as
 \begin{equation}
 \label{eigenstates}
	\left | \psi \right \rangle = \sum_{l,k} \frac{1 }{\sqrt{2}}c_{lk}a_{l}^{\dagger}a_{k}^{\dagger}\left | 0  \right \rangle,
\end{equation} 
with $c_{lk}$ the probability amplitude of two anyons occupying the corresponding Fock state. Note that no restrictions are put on the site labels $l$ and $k$, which means that they can independently go through from $1$ to $L$. In this case, the dimension of the Hilbert space is $L^2$. Substituting Eqs. (\ref{Hamiltonian}) and (\ref{eigenstates}) into the stationary Schr\"odinger equation $H\left | \psi \right \rangle=E\left | \psi \right \rangle$, we obtain the eigenequation with respect to the coefficients $c_{lk}$ as
\begin{eqnarray}
\label{coefficients}
    E c_{lk}=&-J[e^{i\chi \pi(\delta_{lk}+\delta_{l(k-1)})}
	c_{l(k-1)}+e^{-i\chi \pi(\delta_{lk}+\delta_{l(k+1)})}c_{l(k+1)}\notag\\
 &+c_{(l-1)k}+c_{(l+1)k}]+F(l+k)c_{lk},
\end{eqnarray}
which is the very relation Eq.~(4) in Ref.~\cite{Zhang2022}. In this conventional scheme of basis vectors, the probability amplitudes $c_{lk}$ form a matrix as shown in Fig. \ref{fig1}(b), and $c_{lk}$ and $c_{kl}$ are related to each other as follows
 \begin{equation}
 \label{amplitudes relationship}
	c_{lk}=c_{kl}e^{i\chi \pi \epsilon(k-l)}, 
\end{equation} 
due to the commutation relation (\ref{CommutationRelation}). 

Since the Fock states $a_{l}^{\dagger}a_{k}^{\dagger}\left | 0  \right \rangle $ and $a_{k}^{\dagger}a_{l}^{\dagger}\left | 0  \right \rangle $ are linearly dependent, it is not an appropriate way to have simultaneously both of them serving as basis vectors, which are typically linearly independent state vectors and form a complete set for the state space. Thus an alternative and more intuitive choice of the basis vectors are the Fock states with one anyon at site $l$ and the other one at site $k$ of the lattice, but with the restriction of $l\geq k$. It can be easily shown that these new basis vectors, $(1+\delta_{lk})^{-\frac{1}{2}}a_{l}^{\dagger}a_{k}^{\dagger}\left | 0  \right \rangle $ for $l \ge k$, are orthogonal, normal, and complete as required by the quantum theory. So, the dimension of the Hilbert space, i.e., the number of the expansion coefficients, is reduced to $L(L+1)/2$. The eigenstates of the Hamiltonian Eq.~(\ref{Hamiltonian}) are expanded in this orthogonal scheme of basis vectors as 
\begin{equation}
 \label{a3}
    \left | \psi \right \rangle = \sum_{l\ge  k} \frac{1}{\sqrt{1+\delta_{lk}}}c'_{lk} a_{l}^{\dagger}a_{k}^{\dagger}\left | 0  \right \rangle.
\end{equation} 
And, the normalizations of the two-anyon states $|\psi\rangle$ in Eqs.~(\ref{eigenstates}) and (\ref{a3}) require that the coefficients $c$ and $c'$ should satisfy the following relation
\begin{equation}
    \sum_{l,k}|c_{lk}|^2=\sum_{l\ge k}|c'_{lk}|^2=1.
\end{equation}
The eigenequation for the coefficients $c'_{lk}$ is obtained from the stationary Schrödinger equation in just the same way as for Eq. (\ref{coefficients})
\begin{align}
    E c'_{lk}=-J\,\bigg[&\sqrt{1+\delta_{lk}} e^{i\chi \pi \delta_{lk}}c'_{l(k-1)} \nonumber \\ 
    &+(1-\delta_{lk})\sqrt{1+\delta_{l(k+1)}} e^{-i\chi \pi \delta_{l(k+1)}}
	c'_{l(k+1)} \nonumber \\
    &+(1-\delta_{lk})\sqrt{1+\delta_{(l-1)k}}
	c'_{(l-1)k} \nonumber \\
    &+\sqrt{1+\delta_{lk}} c'_{(l+1)k} \bigg]+F(l+k)c'_{lk}. \label{coefficients22}
\end{align}

We illustrate schematically the difference and connection in the structure of two 2D Hilbert spaces, the conventional one with coefficients $c$ and the orthogonal one with coefficients $c'$, as shown in Fig. \ref{fig1}. Clearly, the orthogonal space in the triangular area happens to be the conventional one in the square area folded along the diagonal elements. The eigenstate in both schemes is written in a column vector of coefficients $c$ or $c'$ in an arranged order (see below for examples). The eigenequations for two anyons on a 1D lattice, Eqs. (\ref{coefficients}) and (\ref{coefficients22}), which equivalently describe a single particle jumping between the nearest neighbors on a 2D lattice, have exactly the same form in these two schemes, with the only difference being the phases and delta functions in front of the coefficients. For instance, the probability amplitude with one anyon at site $l$ and the other at site $k$ is indicated by a lattice site $(l, k)$ in 2D. Note that in the conventional basis vector where the lattice sites $l$ and $k$ are unrestricted, the result of this 1D-to-2D mapping covers the entire 2D square lattice. However, in the orthogonal basis vector where they are restricted, i.e. $l \ge k$, one is left with only the lower left triangular region of the 2D square lattice. The hopping occurs only between the nearest-neighbor sites, that is, the coefficients $c_{lk}$ can only jump to its four nearest-neighbor sites. The normal jumps are indicated by ordinary bars, while those with complex phases are denoted by arrows. It is found only those jumps in and out of the diagonal elements along the horizontal $k$ direction, i.e., $(l, l)\rightarrow (l, l\pm 1)$, carry complex phases $e^{\mp i \chi \pi}$ shown as solid/hollow arrows in Fig. \ref{fig1}(a). Note that only one of the two cases of non-zero complex phases in the conventional jumps is preserved in the orthogonal jumps, as can be seen from the remaining Kronecker $\delta$ in the phases in the first two terms of Eq. (\ref{coefficients22}). The site-dependent potential is manifested by the increasing darkening of the lattice site along the diagonal elements, which is the effect of the external force.


 \begin{figure}[t]
	\centering
	\includegraphics[width=1\linewidth]{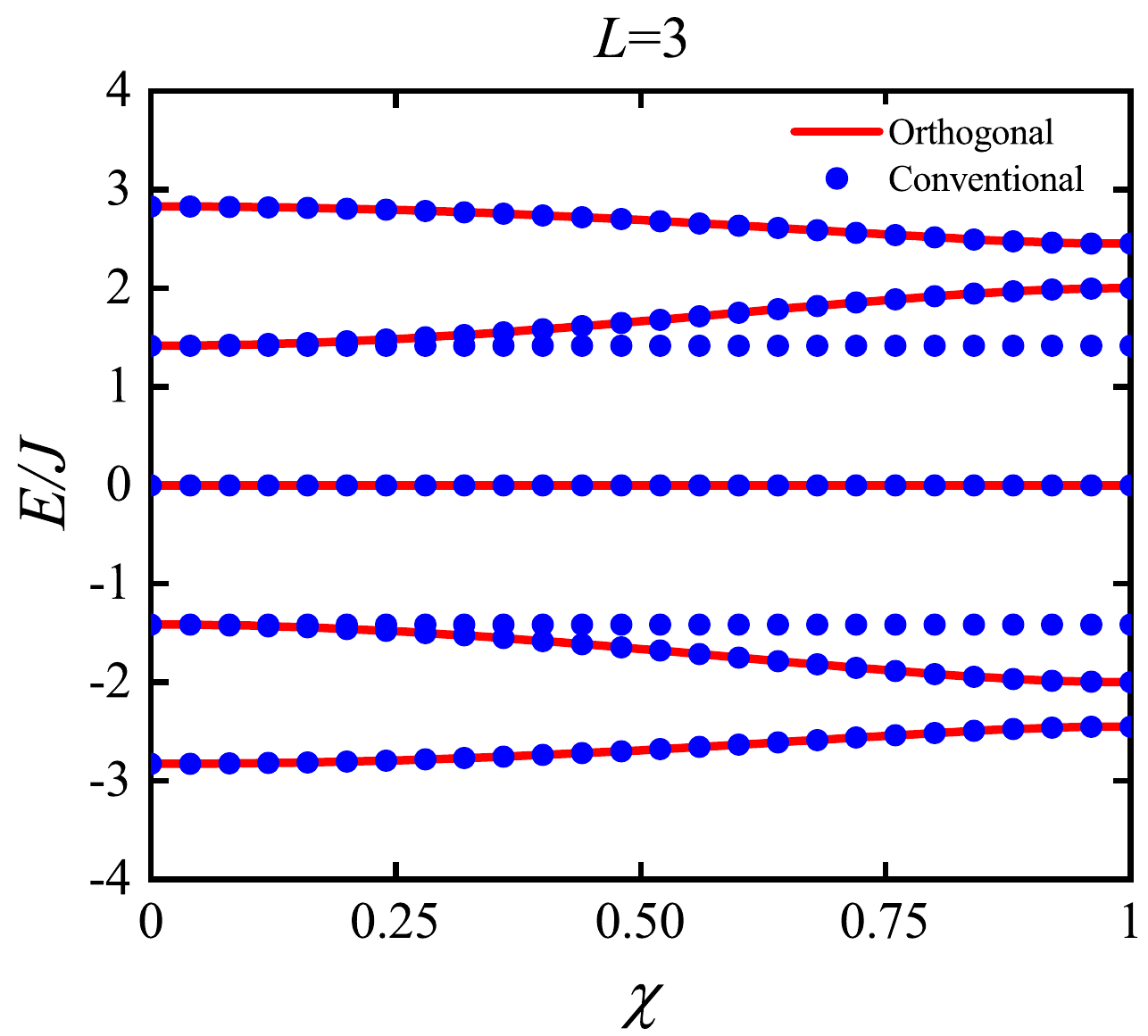}
	\caption{(Color online) The energy spectrum for the two-particle anyon-Hubbard Hamiltonian versus statistical parameter $\chi$ calculated numerically in the conventional Hilbert space (blue dots) and analytical results (\ref{a11}) in the orthogonal space (red lines) with $L=3$ and $F/J=0$. Out of the 9 eigenvalues for each $\chi$ three are non-physical with the coefficients violating the anyonic relation (\ref{amplitudes relationship}). The energies are in units of $J$.} \label{3L}
\end{figure}
\begin{table*}[htbp]
	\centering
 \caption{\small{Eigenvalues and corresponding eigenvectors in conventional and orthogonal Hilbert spaces with $L =3$ and $F=0$ for $\chi=1$.}}
 \label{table}
	\begin{tabular}{p{2cm}<{\centering} p{10cm}<{\centering} p{5.5cm}<{\centering}}
		\specialrule{0.05em}{3pt}{1pt}
		\specialrule{0.05em}{1pt}{5pt}
		\multicolumn{1}{c}{Eigenvalues} & \multicolumn{1}{c}{9-vector $|\psi\rangle $} & \multicolumn{1}{c}{6-vector $|\psi\rangle $ }  \\[0.2cm]
		\specialrule{0.05em}{0.05pt}{6pt}
		$-\sqrt{6}$ & ${(\frac{1}{2 \sqrt{3}},-\frac{1}{2 \sqrt{2}}, 0, \frac{1}{2 \sqrt{2}},-\frac{1}{\sqrt{3}}, \frac{1}{2 \sqrt{2}}, 0,-\frac{1}{2 \sqrt{2}}, \frac{1}{2 \sqrt{3}} )}^T $& ${{(\frac{1}{2 \sqrt{3}}, \frac{1}{2},-\frac{1}{\sqrt{3}}, 0,-\frac{1}{2}, \frac{1}{2 \sqrt{3}})}}^T $ \\[0.15cm]
		$\sqrt{6}$  & ${{(\frac{1}{2 \sqrt{3}}, \frac{1}{2 \sqrt{2}}, 0,-\frac{1}{2 \sqrt{2}},-\frac{1}{\sqrt{3}},-\frac{1}{2 \sqrt{2}}, 0, \frac{1}{2 \sqrt{2}}, \frac{1}{2 \sqrt{3}})}}^T$  & ${(\frac{1}{2 \sqrt{3}},-\frac{1}{2},-\frac{1}{\sqrt{3}}, 0, \frac{1}{2}, \frac{1}{2 \sqrt{3}})}^T$  \\[0.25cm]
		$-2$  & $\frac{1}{2 \sqrt{2}}{(-1, 1, 1,-1, 0, 1,-1,-1, 1)}^T$  & $\frac{1}{2}(-\frac{1}{ \sqrt{2}},-1, 0,-1,-1, \frac{1}{ \sqrt{2}})^T $    \\[0.25cm]
		$2$  & $\frac{1}{2 \sqrt{2}}(-1, -1, 1,1, 0, -1,-1,1, 1)^T $  & $\frac{1}{2}(-\frac{1}{ \sqrt{2}},1, 0,-1,1, \frac{1}{ \sqrt{2}})^T$   \\[0.25cm]
		$0_1$  &$\frac{1}{\sqrt{3}}(1, 0, 0, 0, 1, 0, 0, 0, 1)^T $ & $\frac{1}{\sqrt{3}}(1, 0, 1, 0, 0, 1)^T $  \\[0.2cm]
	$0_{2}$  & $\frac{1}{2}(-1, 0,-1,0,0, 0, 1, 0, 1)^T $  & $\frac{1}{2}(-1, 0,0, \sqrt{2}, 0, 1)^T $   \\[0.2cm]
		  \specialrule{0.05em}{0.05pt}{6pt}$0_3$ & $\frac{1}{2}(0,-1,0,-1, 0, 1, 0, 1, 0)^T $ &    \\[0.25cm]
		$-\sqrt{2}$ &   $\frac{1}{ 2}(0, \frac{1}{ \sqrt{2}}, 1, \frac{1}{ \sqrt{2}}, 0, \frac{1}{ \sqrt{2}}, 1, \frac{1}{ \sqrt{2}}, 0 )^T $  &    \\[0.25cm]
		$\sqrt{2}$ & $\frac{1}{2}(0, \frac{1}{ \sqrt{2}},-1, \frac{1}{ \sqrt{2}}, 0, \frac{1}{ \sqrt{2}},-1, \frac{1}{ \sqrt{2}}, 0)^T $  &     \\[0.25cm]
		\specialrule{0.05em}{4pt}{1pt}
		\specialrule{0.05em}{1pt}{3pt}
	\end{tabular}%
\end{table*}%
 
\section{The structure of the energy spectrum}

The eigenequations (\ref{coefficients}) and (\ref{coefficients22}) are well understood in the language of matrix mechanics by expressing the eigenstates in column vectors. The probability amplitudes $c_{lk}$ are arranged in the column vector in the order of from left to right, from top to bottom, in both schemes. We first take a small-sized system of $L=3$ in the absence of external force $F=0$ as an example, for which the eigenvector and the Hamiltonian in the conventional Hilbert space of dimension $3^2$ read as 
\begin{equation}
|\psi\rangle  = (c_{11},c_{12}, c_{13},c_{21},c_{22}, c_{23},c_{31},c_{32}, c_{33})^T,
\end{equation}
and
\begin{align}
		H_{(L=3)}=-J
		\begin{pmatrix}
			0 & e^{-i\chi\pi} & 0 & 1 & 0 & 0 & 0 &  0& 0\\
			e^{i\chi\pi}& 0 & 1 & 0 & 1 & 0 & 0 & 0 & 0\\
			0 & 1 &  0&  0&  0& 1 &  0&  0& 0\\
			1 & 0 &  0&  0& e^{-i\chi\pi} & 0 &  1& 0 & 0\\
			0 & 1 & 0 &  e^{i\chi\pi}& 0 & e^{-i\chi\pi} &  0& 1 & 0\\
			0& 0 & 1 & 0 &  e^{i\chi\pi}& 0 & 0 &  0& 1\\
			0 & 0 & 0 &  1& 0 & 0 & 0 & 1 & 0\\
			0 &0  & 0 & 0 & 1 &  0&  1& 0 &e^{-i\chi\pi} \\
			0& 0 & 0 & 0 & 0 &  1&  0&  e^{i\chi\pi}&0
		\end{pmatrix},
\end{align}
and in the orthogonal space of dimension $3\times (3+1)/2$ as
\begin{equation}
|\psi\rangle  = (c'_{11},c'_{21},c'_{22}, c'_{31},c'_{32}, c'_{33})^T,     
\end{equation}
and 
\begin{equation} 
	\label{a10}
	H^{\prime}_{(L=3)}=-J
	\begin{pmatrix}
		0&  \sqrt{2} &  0&  0&  0& 0\\
		\sqrt{2}&  0&  \sqrt{2}e^{-i\chi\pi}&  1&  0&0 \\
		0&  \sqrt{2}e^{i\chi\pi}&  0&  0&  \sqrt{2}& 0\\
		0& 1&  0&  0&  1&0 \\
		0&  0&  \sqrt{2}&  1&  0 & \sqrt{2}e^{-i\chi\pi}\\
		0&  0&  0&  0&  \sqrt{2}e^{i\chi\pi}&0
	\end{pmatrix},
\end{equation}
respectively. Note that the Hamiltonian matrices in both spaces are Hermitian which assures the reality of the eigenvalues and the orthogonality of the eigenvectors. While the high-order algebraic equation hinders us from providing an analytical expression for the eigenvalues of a $9\times 9$ matrix, luckily the Hamiltonian in the orthogonal space is analytically solvable in the absence of external force and the 6 eigenvalues for arbitrary $\chi$ are given by
\begin{align}
\label{a11}
	&E_{1,2}=-\sqrt{5\pm \sqrt{5+4 \cos \chi \pi} }, \\
	&E_{3,4}=0, \\ 
	&E_{5,6}= \sqrt{5\mp \sqrt{5+4 \cos \chi \pi} },
\end{align}
each pair of which is positive, zero, and positive, respectively. We compare these solutions with the numerically obtained eigenvalues in the 9-dimensional space in Fig. \ref{3L}. Three additional eigenvalues emerge in the conventional scheme of the eigenequation, which are all $\chi$-independent as illustrated by the blue horizontal lines. The zero eigenvalue is degenerate with the other two zero eigenvalues ($E_{3,4}$) in the whole range of $\chi$, and the other two merge into $E_2$ and $E_5$ in the bosonic limit $\chi=0$. To understand more clearly the nature of these statistics-independent eigenstates, we pay attention to the eigenvalues and their corresponding eigenvectors in the pseudo-Fermion limit $\chi=1$, which are listed in Table \ref{table}. For the physical states with eigenvalues $\pm \sqrt{6}, \pm 2, 0_1, 0_2$, the coefficients in the 9-vector $\left|\psi\right\rangle$ satisfy the relation required by the anyonic commutation relation, that is $c_{lk}=-c_{kl}$ from Eq. (\ref{amplitudes relationship}). \begin{figure}[t]
\centering
\includegraphics[width=1\linewidth]{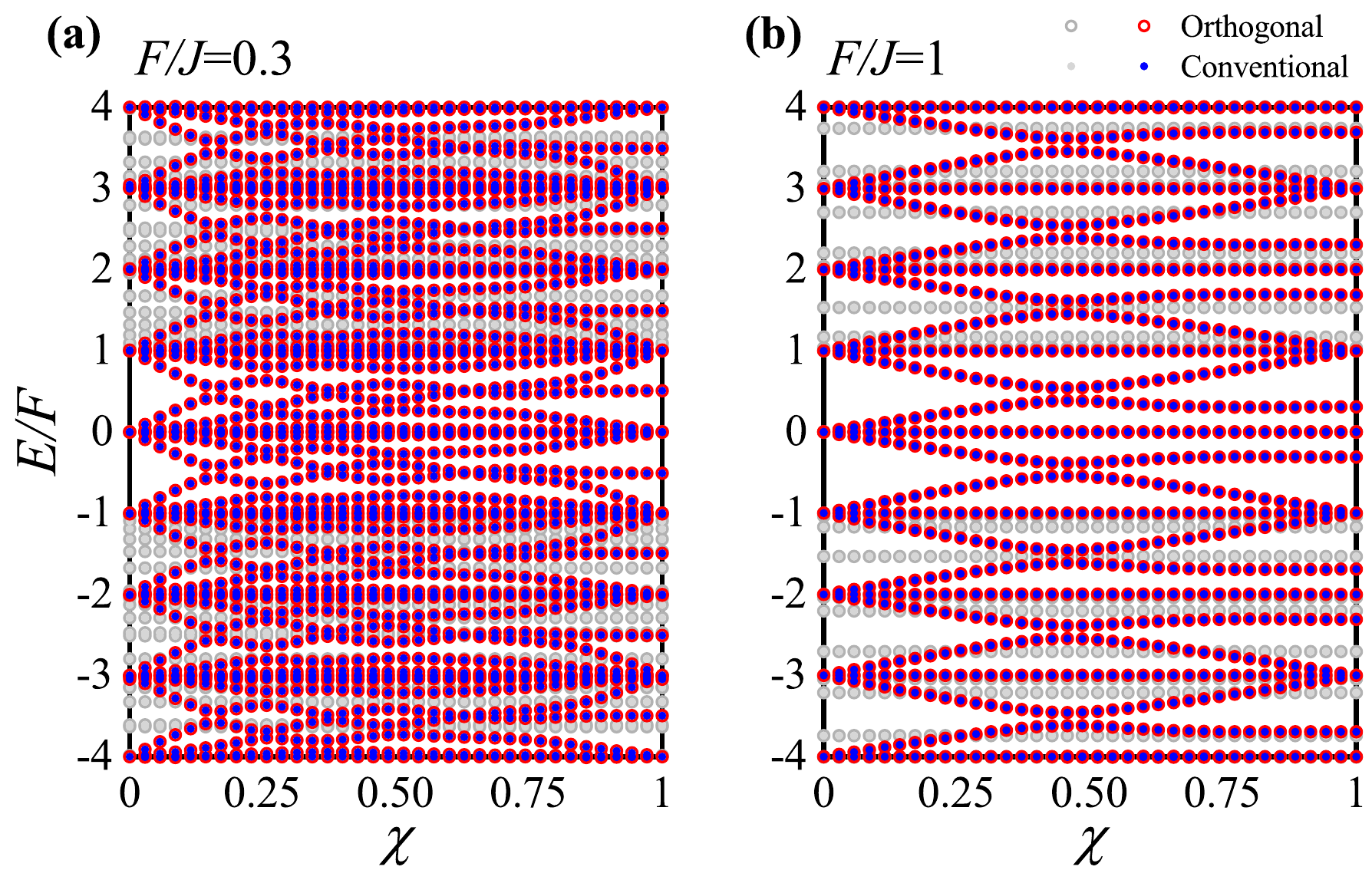}
\caption{(Color online) Comparison of the energy spectrum of the two-anyon lattice system for $L=41$ versus the statistical parameter $\chi$ for (a) $F/J = 0.3$, and (b) $F/J = 1$, numerically computed from the two choices of basis vectors. For clarity the energies in the range $(-4F,4F)$ are shown in units of $F$. To avoid finite-size effects, we kept only the eigenvalues whose eigenvectors showed the largest overlap with the center of the lattice, where the grey dots are the energy values of this criterion that were screened out.}\label{energy spectrum middle}
\end{figure}
For instance, the second and the fourth amplitudes are opposite to each other, i.e. $c_{12}=-c_{21}$. The relation still holds for the two vectors for zero eigenvalues $0_1$ and $0_2$, in which case $c_{12}=-c_{21}=0$. This correctly related coefficients pair assures that we can fold the 9-vectors into 6-vectors in the Hilbert space, as listed in the last column in Table \ref{table}. However, the other three eigenstates in the last three rows in Table \ref{table} violate this basic coefficients relation. We always find that $c_{lk}=+c_{kl}$, i.e. the second and the fourth amplitudes take equal values of $-1, 1/\sqrt{2},1/\sqrt{2}$ in the 9-vectors for eigenvalues $0_3, -\sqrt{2}, \sqrt{2}$, respectively. This makes the folding of the 9-vectors into 6-vectors impossible. We conclude that these statistics-independent eigenstates which violate the basic anyon coefficients relation are non-physical and redundant and should be excluded from the two-anyon Hilbert space in the related calculations.

\begin{figure}[t]	
	\centering	\includegraphics[width=1\linewidth]{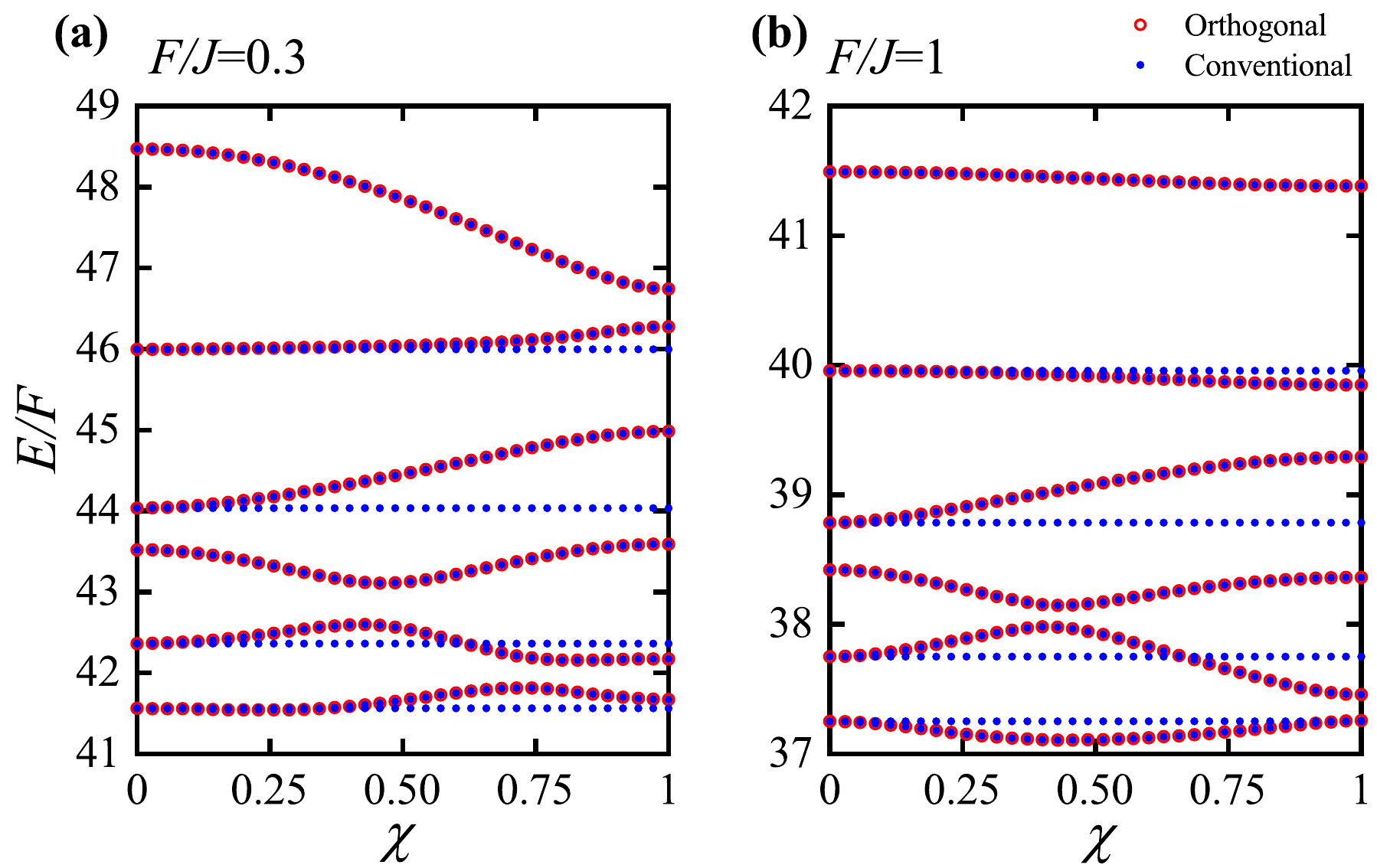}
	\caption{(Color online) The upper edge of the energy spectrum for $L=41$ versus the statistical parameter $\chi$ with external forces (a) $F/J = 0.3$, and (b) $F/J = 1$.  } 
 \label{energy spectrum edge}
\end{figure}

To see how these non-physical eigenstates would change the energy spectrum structure of the two-anyon system, we consider a finite lattice size of $L=41$ and numerically compute the eigenvalues and corresponding eigenvectors in the orthogonal space and compare them with those in the conventional space, as shown in Figs. \ref{energy spectrum middle} and \ref{energy spectrum edge}. This comparison is crucial in explaining the Bloch oscillation period of pseudo-fermions and clarifying the roles of $\chi$ and $F$ in the onset of BOs in the probability densities and correlation functions. The external force leads to the on-site potential energy and we take two force parameters $F/J=0.3$ and $F/J=1$ as examples. To avoid the finite-size effects, let us first pay attention to the middle energy range $(-4F,4F)$. For a better visual effect, we make some eigenvalue selections, and the eigenvalues are plotted in two different styles. Those whose eigenvectors show the largest overlap with the center of the lattice are plotted in blue solid dots in the conventional space and red open circles in the orthogonal spaces, respectively, i.e. we choose the largest coefficient in each eigenvector in such a way that both the distance between any two particles at sites $l$ and $k$ and that between each particle and the central lattice site are within a certain range $\Delta r$. Those out of this range are plotted in light grey dots and dark grey open circles accordingly. The equally spaced energy spectral structures as signs of Bloch oscillation can be manifested by choosing the appropriate range $\Delta r$ for different force parameters $F$, e.g. $\Delta r=14$ for $F=0.3$ and $\Delta r=8$ for $F=1$, respectively, in Figs. \ref{energy spectrum middle}(a) and \ref{energy spectrum middle}(b). Interestingly we adopt the same selection rules for the spectrum in the conventional and orthogonal schemes and find no difference in the middle energy range, i.e. in both schemes there exists this evenly spaced structure in the spectrum for $\chi=1$ and $F/J=0.3$ with spacing $F/2$ rather than $F$. This suggests that the orthogonality of the basis vectors does not affect the BOs phenomenon for a large enough lattice size. Note that there exists a large amount of redundant eigenstates in the middle energy range such as that with eigenvalue $0_3$ in the $L=3$ case which are all degenerate with some specific normal states, e.g. those with eigenvalues $0_1,0_2$. These redundant non-physical states hidden in the high-fold degeneracy, however, shall be excluded in the calculation for dynamical evolution in the next section. On the other hand, when we carried out the calculation in the full energy spectrum, a significant difference occurs at the edge area of the spectrum between the two schemes, where a large amount of non-physical eigenvalues from the conventional diagonalization scheme persist with the anyonic coefficients relation broken in the eigenvectors. We show in Fig. \ref{energy spectrum edge} the numerical results of the edge range of energy spectrum for $L=41$ with $F/J=0.3$ and $F/J=1$. Note that the selection rule has not been applied to the energy levels as the BOs phenomenon does not appear in the upper or lower edges of the spectrum.

 \begin{figure}[t]	
 	\centering
 	\includegraphics[trim=5 0 0 0,clip,width=1\linewidth]{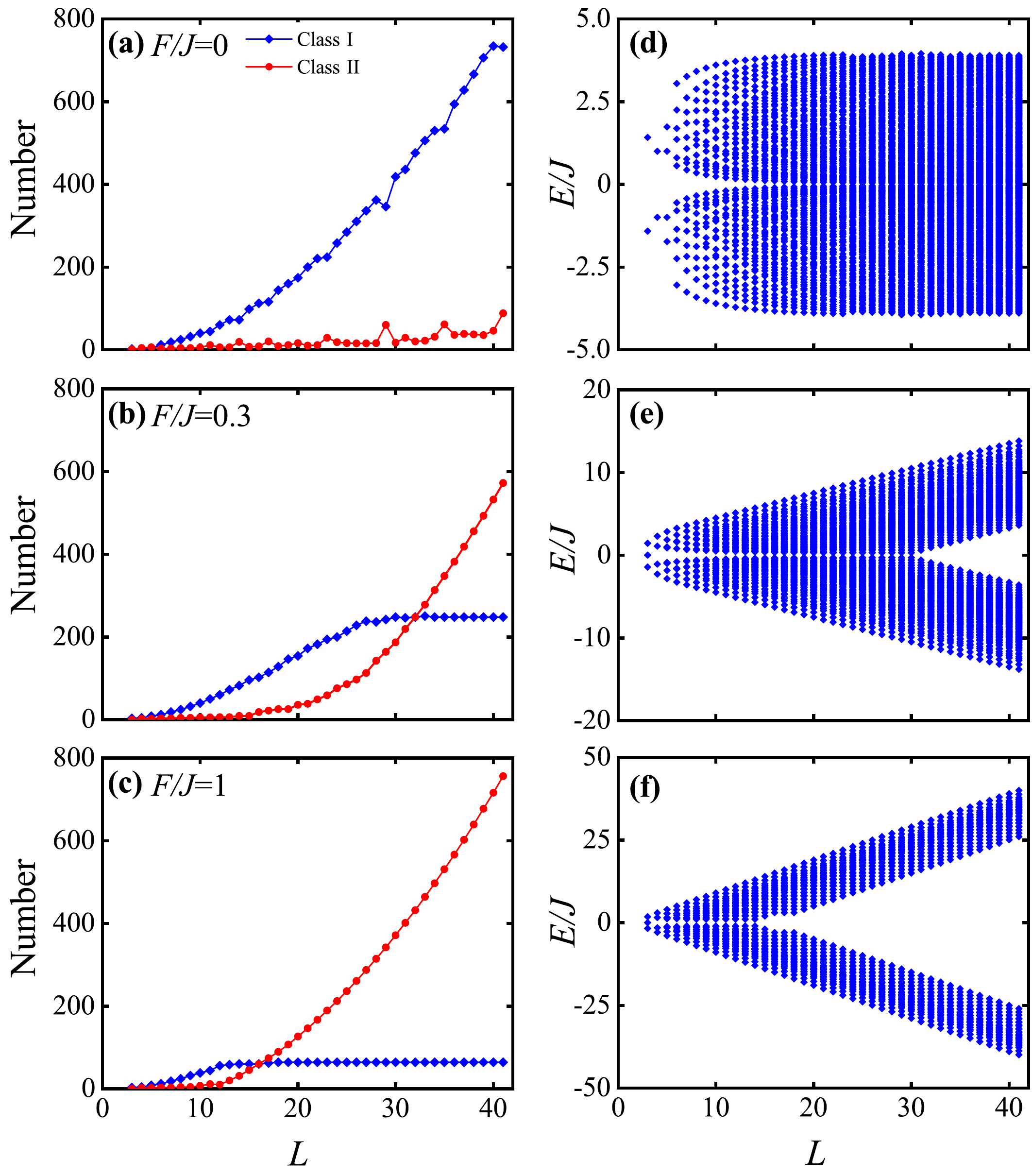}
 	\caption{(Color online) The number of redundant eigenvalues and the distribution of Class I redundant eigenvalues with increasing lattice size $L$ for $\chi=1$ with external forces $F/J = 0, 0.3$ and $1$ from top to bottom. } \label{redundant}
 \end{figure}

One may wonder how many of these redundant non-physical states are and how the number of these states increases with the lattice size. It is possible to classify these redundant states into two classes, those never appear in our orthogonal scheme (Class I, denoted by pure blue dots in the spectrum), and those are degenerate with the eigenvalues in the orthogonal scheme (Class II, denoted by blue dots and red circles overlapping lines). For instance, in the case of $L=3$, the eigenvalue $0_3$ of $E=0$ belongs to Class II, while eigenvalues $\pm \sqrt{2}$ are Class I. We consider again the pseudo-fermion statistical limit of $\chi=1$ in which case it is easy to identify whether the redundant eigenvalues are degenerate with those of the physical states. By checking the opposite relation of any pair of coefficients on transposed positions, say $c_{12}$ and $c_{21}$, the redundant eigenvalues whose coefficients violate this relation are shown with the increasing number of lattice size in Fig. \ref{redundant} for external forces $F/J=0,0.3,1$, respectively. For clarity, we only show the Class I redundant eigenvalues on the right column and count the numbers of Class I and Class II eigenvalues on the left. Clearly, the larger lattice hosts more redundant eigenvalues according to $L(L-1)/2$ and the ratio of redundant to orthogonal eigenvalues increases with the number of lattice sites $L$ as $(L-1)/(L+1)$ approaching unity. We observed that the distribution of these redundant eigenvalues is significantly influenced by the external force $F$. For $F=0$ almost all redundant eigenvalues belong to Class I and are distributed evenly in the whole spectrum. As $F$ increases, the Class I redundant states are expelled out of the middle energy range for larger lattice sizes. Their number tends to saturate to a lattice-size independent value and Class II eigenvalues shall dominate. These redundant eigenvalues will take part in the dynamical evolution of the system when one starts from an initial state whose coefficients do not meet the requirement for anyons as discussed in the next section.

\section{The Dynamical behavior}   

The nonlocal feature of anyons may exhibit many-body effects in the time evolution of the system even in the absence of on-site interaction. Now we turn to the dynamics of the two correlated anyons in a 1D lattice and explore the time evolution of several observables such as the revival probability, the probability density, and the correlation functions. With this aim, let us first introduce the time-dependent probability amplitude $c_{lk}(t)$ of finding one anyon at site $l$ and the other one at site $k$ of the lattice at time $t$. The expansion of any state $\left | \psi(t) \right \rangle$ upon the basis vector takes the same form as Eqs.~(\ref{eigenstates}) and (\ref{a3}) except that the coefficients $c_{lk}(t)$ are now time-dependent. The evolution equations for the amplitudes $c_{lk}(t)$, as obtained from the Schr\"odinger equation $i\partial_t \left | \psi (t) \right \rangle =H \left | \psi(t)  \right \rangle $ with $\hbar =1$, have the same form as Eqs.~(\ref{coefficients}) and (\ref{coefficients22}) by simply replacing the stationary terms on the left-hand side $E c_{lk}$ by the time derivative terms $i\partial_t c_{lk}$ or $i\partial_t c'_{lk}$, respectively.

We first introduce the revival probability, or the Loschmidt echo, which quantifies the probability of the time-evolved state in the initial configuration
\begin{equation}
\label{a14}
	P_{rev}(t)={|\langle \psi(0)|\psi (t) \rangle|}^{2},
\end{equation} 
The probability density function measures the density distribution on each site of the 1D lattice 
\begin{equation}
\label{a15}
 	 P_{q}(t)= \frac{1}{2}\langle\psi(t)|n_{q}|\psi (t) \rangle, 
\end{equation} 
for $q =1,2,...,L$. Here the factor $1/2$ assures the total probability normalized to unity in the whole lattice. The correlation between the two anyons is described by the two-body correlation  function in position space, 
\begin{equation}
\label{a16}
\Gamma_{qr}(t)= \langle\psi(t)|a_{q}^{\dagger}a_{r}^{\dagger}a_{r}a_{q}|\psi (t) \rangle, 
\end{equation}
with $q, r =1,2,...,L$. The matrix element of which $\Gamma_{qr}(t)$ represents the probability of detecting one particle at site $q$ and its twin particle at site $r$. In the above definitions, the initial state $|\psi (0) \rangle$ is represented in a column vector of $c_{lk}(0)$ or $c'_{lk}(0)$, the state vector $|\psi (t) \rangle= e^{-iHt}|\psi (0) \rangle$ at time $t$ is expressed as a column vector of coefficients $c_{lk}(t)$ or $c'_{lk}(t)$ (in the order of from left to right, from top to bottom, similar to the $L=3$ case), and $H$ is written as a matrix $H_{(L)}$ or $H'_{(L)}$ in the conventional space with dimension $L^2$ or in the orthogonal space with dimension $L(L+1)/2$, respectively. 

These dynamical quantities can be calculated directly from the coefficients $c_{lk}$ in the conventional scheme with results
\begin{equation}
 	 P_{rev}(t)= {\left| \sum_{l, k}c_{lk}(0)c_{lk}(t) \right|}^2, 
\end{equation}
for the revival probability,
\begin{align}
  P_{q}(t)= \frac{1}{4}\sum_{l}\bigg\{&{|c_{lq}(t)|}^2+{|c_{ql}(t)|}^2 \\ \nonumber
  &+\big[c^{\ast}_{lq}(t)c_{ql}(t)e^{-i\chi\pi\epsilon(l-q)}+h.c.\big]\bigg\},   
\end{align}  
for the probability density function, and
\begin{align}
  \Gamma_{qr}(t)= \frac{1}{2}\bigg\{&{|c_{rq}(t)|}^2+{|c_{qr}(t)|}^2\\ \notag
   &+\big[c^{\ast}_{qr}(t)c_{rq}(t)e^{-i\chi\pi\epsilon(q-r)}+h.c.\big]\bigg\} , 
\end{align}
for the correlation function, respectively. In the orthogonal scheme, one has instead
\begin{equation}
 	P_{rev}(t)= {\left|\sum_{l\ge k} c'_{lk}(0)c'_{lk}(t) \right|}^2 , 
\end{equation} 
for the revival probability, \begin{equation}
 	 P_{q}(t)= \frac{1}{2} \bigg[\sum_{l\ge q}{| c'_{lq}(t)|}^2+\sum_{l\le q}{| c'_{ql}(t)|}^2\bigg], 
\end{equation} 
for the probability density function, and
\begin{equation}
\Gamma_{qr}(t)=
\left\{ 
    \begin{array}{lc}
        | c'_{rq}(t)|^2 & r>q, \\
        2| c'_{qq}(t)|^2 & r=q, \\
        | c'_{qr}(t)|^2&r<q.\\
    \end{array}
\right. 
\end{equation}
for the correlation function.

\begin{figure}[t]		
 	\centering
 	\includegraphics[trim=5 0 0 0,clip,width=1\linewidth]{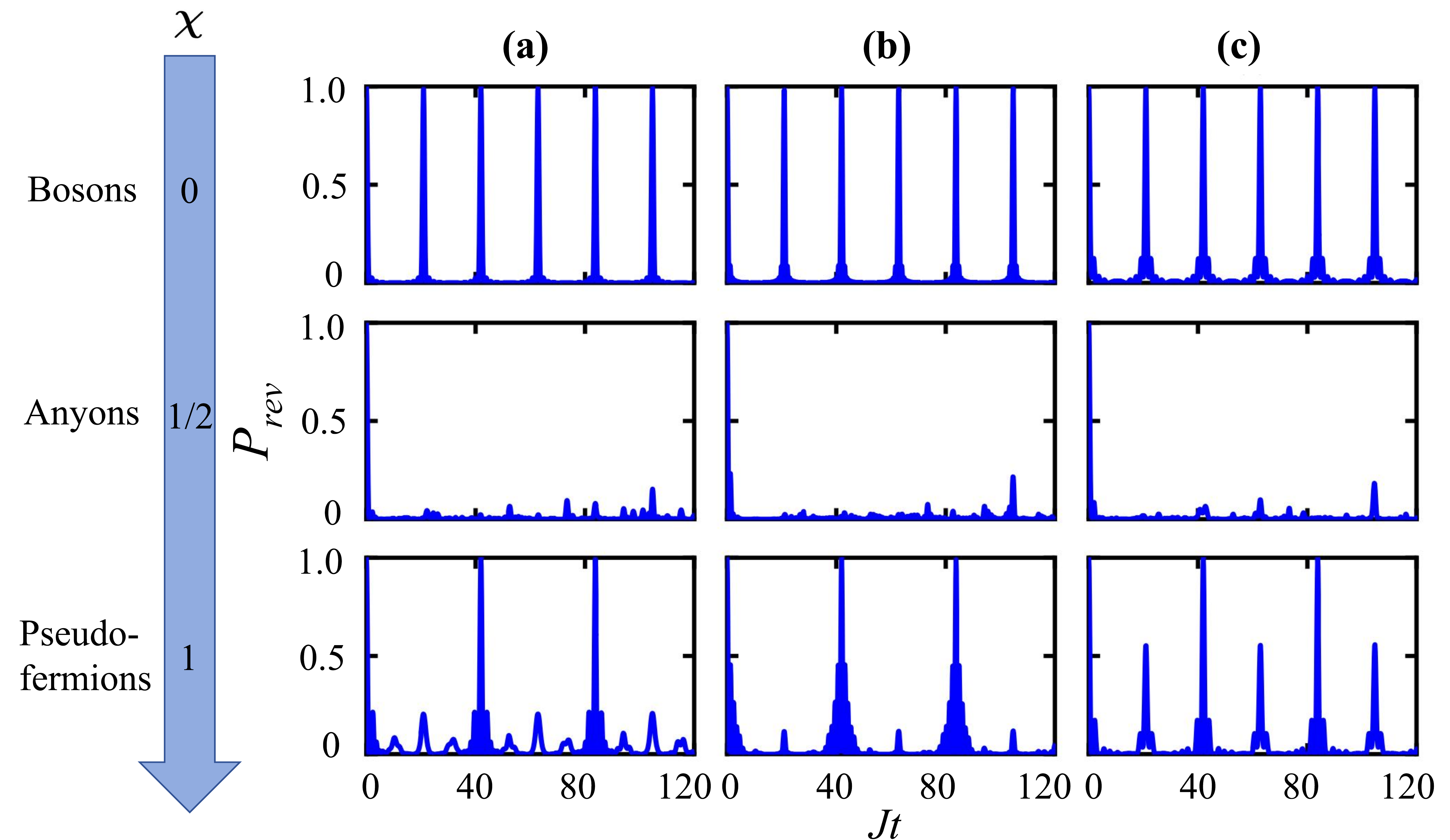}
 	\caption{(Color online) Evolution of the revival probability of two correlated anyons undergoing BOs on a 1D lattice for $F/J = 0.3$, $L = 41$ and for increasing values of the statistical phase exchange $\chi = 0, 0.5$ and $1 $ from top to bottom ($\chi = 0$ corresponds to non-interacting bosons, whereas $\chi = 1 $ corresponds to pseudofermions). The results calculated in the conventional and orthogonal spaces are exactly the same and we show the revival probability evolution for three initial states, i.e. $\left|\phi_{1}\right\rangle$ with two anyons on the same site (a), $\left|\phi_{2}\right\rangle$ on the neighboring sites (b), $\left|\phi_{3}\right\rangle$ and on the next-neighboring sites with an empty site in between (c), respectively.} \label{Prev}
\end{figure}

For the study of quantum walking, the two anyons are initially prepared in the same site, often located in the center of the lattice, and observe the dynamics thereafter. To check the inadequacy of the conventional choice of the Hilbert space, here we discuss three distinct initial states for the simulation of system dynamics, i.e., two anyons on the same site, on the neighboring sites, and on the next-neighboring sites with an empty site in between, denoted by red, yellow, and green stars in Fig. \ref{fig1}. We see that these initial states in the 2D lattice are on, one-step away, and two-step away from the diagonal line, respectively. (i) The mostly studied initial state with coefficients $c_{lk}(0)=\delta_{l,c}\delta_{k,c}$, i.e. $\left|\phi_{1}\right\rangle \equiv (1/\sqrt{2})a_c^{\dagger 2}\left|0\right\rangle $, is expressed in our notation as a column vector 
\begin{equation}
\left|\phi_{1}\right\rangle:\qquad(0,... 1_{c,c},...0)^T
\end{equation}
of dimension $L^2$ in conventional Hilbert space and a vector in the same form yet of dimension $L\times (L+1)/2$ in orthogonal scheme, with the subscript indicating that the two anyons are equally prepared in the site $c$. This arrangement sheds light on how the system behaves when the particles are closely packed together. (ii) The second initial state with $c_{lk}(0)=\delta_{l,(c+1)}\delta_{k,c}$, i.e. $\left|\phi_{2}\right\rangle \equiv a^\dagger_{c+1}a^\dagger_{c}\left|0\right\rangle $, is represented as a vector \begin{equation}
\label{b}\left|\phi_{2}\right\rangle:\qquad \frac{1}{\sqrt{2}}(0,...1_{c,c+1}e^{i\chi\pi},...1_{c+1,c}...0)^T
\end{equation}
in the conventional space and $(0,...1_{c+1,c}...0)^T$ in the orthogonal space with the same dimensions as in initial state $\left|\phi_{1}\right\rangle$, signifying that two particles are positioned at the nearest neighbors in the lattice center. (iii) The third state  with $c_{lk}(0)=\delta_{l,(c+1)}\delta_{k,(c-1)}$, i.e. $\left|\phi_{3}\right\rangle \equiv a^\dagger_{c+1}a^\dagger_{c-1}\left|0\right\rangle $, is represented in a similar way as a vector
\begin{equation}
\left|\phi_{3}\right\rangle:
\qquad \frac{1}{\sqrt{2}}(0,...1_{c-1,c+1}e^{i\chi\pi},...1_{c+1,c-1}...0)^T
\end{equation} 
in the conventional space and $(0,...1_{c+1,c-1}...0)^T$ in the orthogonal space, indicating that two particles are in the next nearest neighboring sites. Note that in the representation of the initial states $\left|\phi_{2}\right\rangle$ and $\left|\phi_{3}\right\rangle$, an appropriate phase factor is included in the transposed position of the non-zero coefficients at position $(c+1,c)$ or $(c+1,c-1)$, which assures that they describe a possible state function of two anyons. It is easy to check that the projection of this properly represented state onto any one of the non-physical eigenstates is zero. In this sense, the initial state $\left|\phi_{1}\right\rangle$ is trivial as the overlapping of its representation in conventional scheme with the non-physical states is already zero (see the $L=3$ case for an example: the only non-zero element in the state $\left|\phi_{1}\right\rangle$ is the fifth, while the fifth elements of all three non-physical states are zero).  On the other hand, in the simulation of the two-anyon Hamiltonian by means of the electric circuits, the wavefunction satisfying the Schr\"odinger equation may not carry this phase factor, which leads to a state
\begin{equation}
\label{b'} \left|\phi_2'\right\rangle:\qquad \frac{1}{\sqrt{2}}(0,...1_{c,c+1},...1_{c+1,c}...0)^T
\end{equation}
or it is even possible to prepare a state like
\begin{equation}
\label{b''} \left|\phi_2''\right\rangle:\qquad (0,...0_{c,c+1},...1_{c+1,c}...0)^T
\end{equation}
These initial states, however, do not describe two-anyon states, and their expansions in the conventional Hilbert space inevitably involve the non-physical states, which leads to distinct dynamical results. For a lattice size $L=41$ adopted in the following numerical simulation, the lattice center is located at $c = 21$.
\begin{figure}[t]	
 	\centering
 	\includegraphics[trim=5 0 0 0,clip,width=1\linewidth]{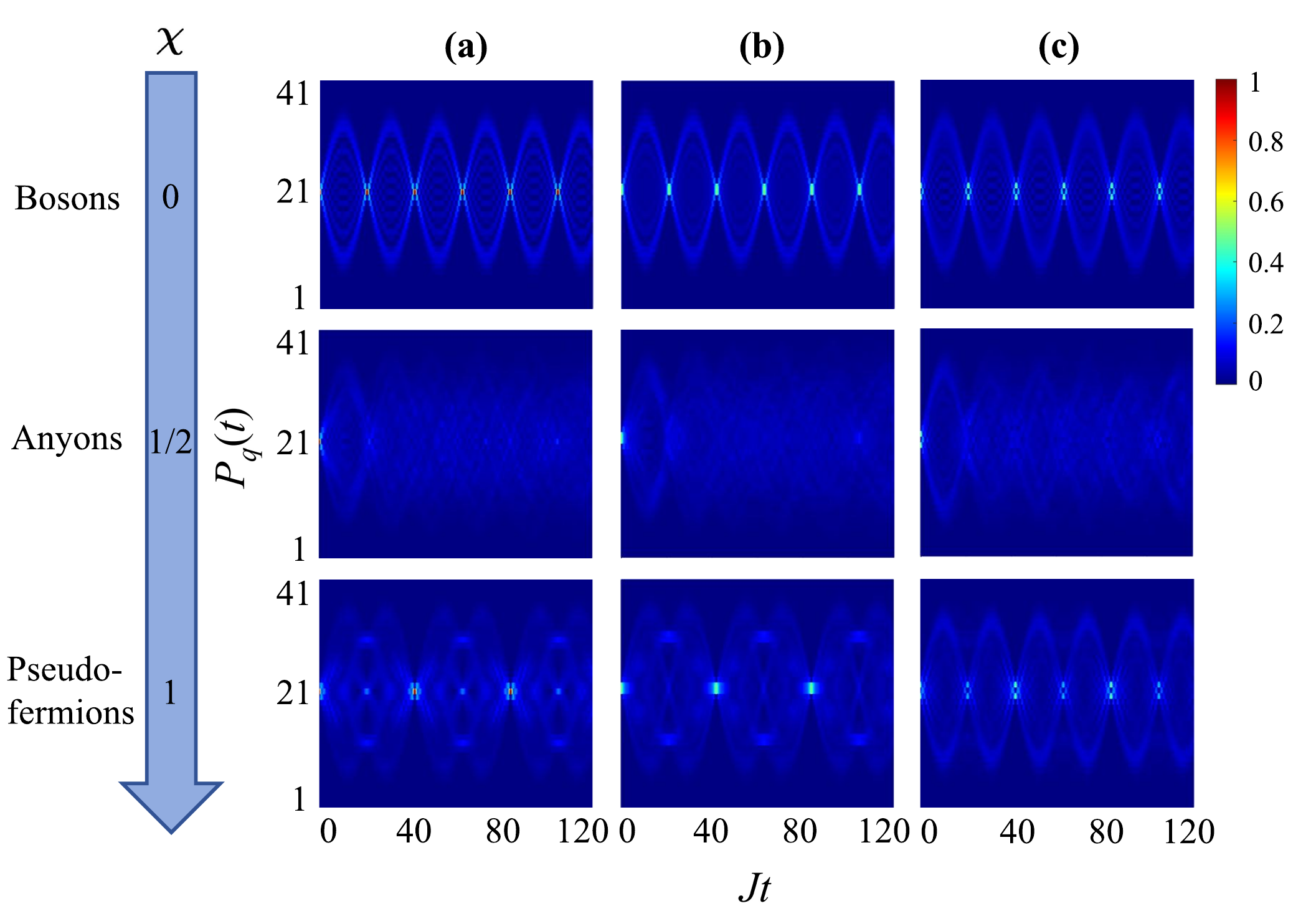}
	\caption{(Color online) Evolution of the probability density function of two correlated anyons undergoing BOs on a 1D lattice for $F/J = 0.3$, $L = 41$ and for increasing values of the statistical phase exchange $\chi = 0, 0.5$ and $1 $ from top to bottom. The results calculated in the conventional and orthogonal spaces are exactly the same and we show the probability density evolution for three initial states, i.e. $\left|\phi_{1}\right\rangle$ with two anyons on the same site (a), $\left|\phi_{2}\right\rangle$ on the neighboring sites (b), $\left|\phi_{3}\right\rangle$ and on the next-neighboring sites with an empty site in between (c), respectively.} 
 \label{Pk}
\end{figure} 

The dynamics of the system depends crucially on the initial states. With properly represented states $\phi_i$ ($i=1,2,3$) with the phase factor $e^{i \chi \pi}$ included in correct positions we observe exactly the same dynamical behavior in either the conventional or orthogonal schemes. The revival probability for initial sate $\left|\phi_{1}\right\rangle$ is easily calculated as $P_{rev}(t)=|c_{cc}(t)|^2=|c'_{cc}(t)|^2$ in the two schemes, which exhibits the degraded Bloch oscillation behavior \cite{Longhi2012} for statistical parameter $\chi$ away from the bosonic and pseudo-fermionic limits as shown in Fig. \ref{Prev}(a). When the two anyons in the initial state are not located on the same lattice site, we have $P_{rev}=2|c_{c+1,c}(t)|^2=|c'_{c+1,c}(t)|^2$ and $P_{rev}=2|c_{c+1,c-1}(t)|^2=|c'_{c+1,c-1}(t)|^2$ for states $\left|\phi_2\right\rangle$ or $\left|\phi_3\right\rangle$ respectively, the results of which show the similar double period phenomena of Bloch oscillation, that is, the period of BOs in the pseudo-fermionic limit is twice of that in the bosonic limit $T_B=2\pi/F$, while the oscillations disappear for statistical parameters in between, for instance, at $\chi=1/2$, as shown in Figs. \ref{Prev}(b) and (c). The coefficients at time $t$ still satisfy the anyonic relation such that the off-diagonal probability in the orthogonal space is twice the value in conventional space, while the diagonal ones are equal. For state $\left|\phi_3\right\rangle$, the only difference that occurs at the revival probability is that the secondary peaks tend to be more prominent. We find this happens equally for other initial states with further increasing distance between the two anyons in the initial state, i.e. the revival probability may reach as high as $0.7$ at the original period times of the BOs. 

The absence of BOs in the case of statistical parameter $0<\chi<1$ can be more clearly seen in the evolution of the probability density function $P_q(t)$ starting from different initial states introduced above. The two anyons will return to their original positions periodically when the BOs occur in the bosonic limit, while it takes twice the time in the pseudo-fermionic limit, no matter when they are put in the same site ($\phi_1$) or different sites ($\phi_2$) and ($\phi_3$), see the top and bottom panels of Fig. \ref{Pk}. Away from these two limiting cases, for instance, $\chi = 1/2$, the anyons would never return to their initial positions for all three initial states. For the initial state $\left|\phi_{3}\right\rangle$ with an empty site in between, in the pseudo-fermion limit we still find a very large probability for the two anyons to return to their original positions at the ordinary period, which is consistent with the period in the revival probability in Fig. \ref{Prev}. We find this happens for other initial states when the anyons are separated by more empty sites.

\begin{figure}[t]	
 	\centering
 	\includegraphics[trim=5 0 0 0,clip,width=1\linewidth]{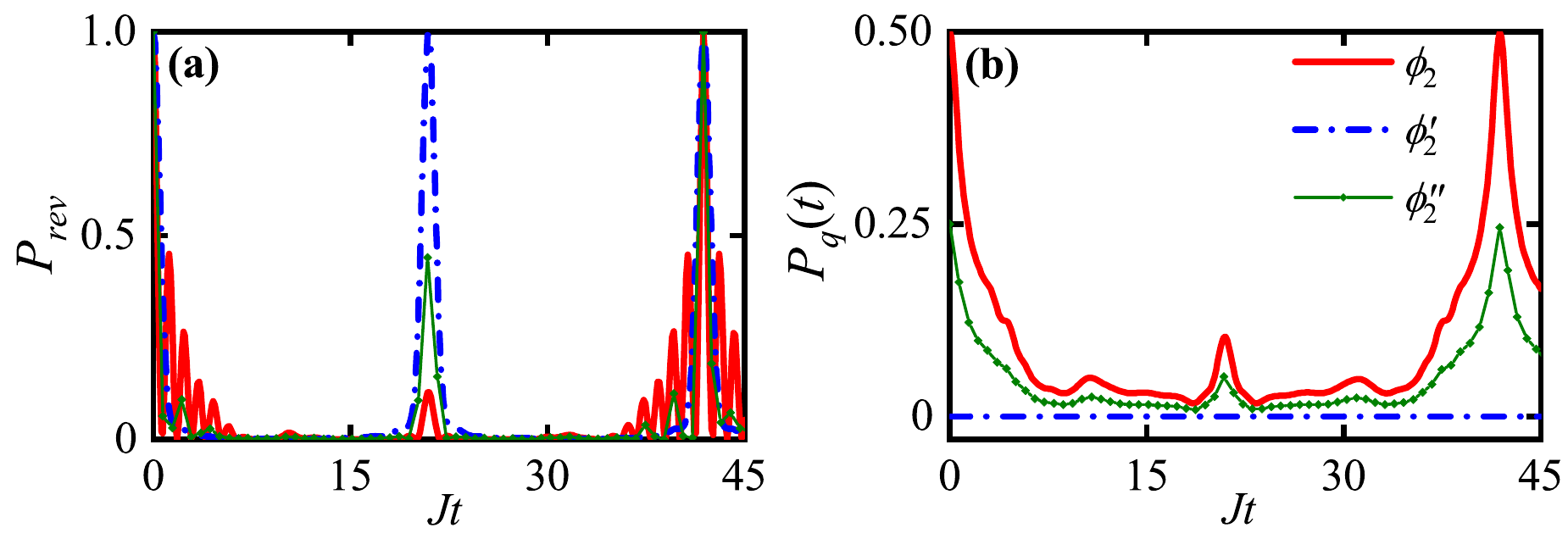}
 	\caption{(Color online)(a) The dynamics of revival probability and (b) probability density at the central site (here $c=21$) from the initial states $\left|\phi_{2}\right\rangle$, $\left|\phi'_{2}\right\rangle$ and $\left|\phi''_{2}\right\rangle$ with $F/J = 0.3$ and $L = 41$ in the fermionic limit $\chi = 1$.} 
  \label{non-anyonic}
\end{figure}

It is necessary to examine carefully the details of the dynamics of those "non-anyonic" initial states. For the statistical parameter $0 < \chi < 1$, the newly calculated results in the conventional Hilbert space show drastically different behavior for the three states. Specifically, we have $P_{rev}=(1/2)|c_{c,c+1}(t)+c_{c+1,c}(t)|^2$ and $P_{rev}=|c_{c+1,c}(t)|^2$ for states $\left|\phi'_{2}\right\rangle$ and $\left|\phi''_{2}\right\rangle$, respectively. Note that the coefficients here do not satisfy the relation (\ref{amplitudes relationship}), and it is not possible to further simplify this probability. To see more clearly the difference with their anyonic partner state $\left|\phi_{2}\right\rangle$, we illustrate the revival probability of these states in Fig. \ref{non-anyonic}(a) for $F/J=0.3$ and in the pseudo-fermionic limit where the most significant discrepancy occurs. It can be seen that all states experience a full revival at $2T_B$ and the anyonic state shows only 0.1 revival probability at the time of BOs period $T_B$. This means the wave function has little chance of returning to the initial state at time $T_B$ when the two anyons are prepared in different sites initially. The non-anyon states $\phi'_2$, on the other hand, get a $100\% $ revival probability at this specific moment $T_B$, and even the state $\left|\phi''_2\right\rangle$ will get a $50\%$ revival probability. This conclusion holds equally for initial state $\left|\phi_{3}\right\rangle$ where the two anyons are separated by an empty site initially. 

The probability density at the central site demonstrates even more discernible features for the experiment simulation. It is shown that while the robust BOs at $2T_B$ still occur in the central site density for anyonic initial state $\left|\phi_{2}\right\rangle$, that for non-anyonic initial state $\left|\phi'_{2}\right\rangle$ vanishes completely in the entire dynamical evolution, as shown by the constant zero blue-dot-dashed line in Fig. \ref{non-anyonic}(b). This is easy to understand that the probability density (\ref{a15}) is defined to measure the number of anyons in the lattice, which is the expectation value of the number operator $n_q$ on the state $\phi(t)$ at time $t$. Note that the number operator even does not take a diagonal form in the conventional space, as the Fock states basis $a_{l}^{\dagger}a_{k}^{\dagger}\left | 0  \right \rangle $ and $a_{k}^{\dagger}a_{l}^{\dagger}\left | 0  \right \rangle $ equally contribute to the expectation value of number operator $n_q$ at $q=k$. When we design the initial state in a way that the coefficients satisfy the anyonic relation such as $\left|\phi_{2}\right\rangle$, the measurement of the probability at the central site $c=21$ gives $1/2$ as the two anyons are initially located at $c=21$ and $c+1=22$, which may oscillate as the anyons walk to other lattice sites. Starting from a non-anyonic state such as $\left|\phi'_{2}\right\rangle$, we have no chance to find any anyon in the system at $t=0$ as well as in the later time $t>0$. The state $\left|\phi''_{2}\right\rangle$ is a mixture of the pure anyonic and non-anyonic initial state, which proves to be a superposition of $\left|\phi_{2}\right\rangle$ and $\left|\phi'_{2}\right\rangle$, leaving only half of the probability for site $q$ or for the whole lattice. In this sense, the simulation of these wave functions does not describe the behavior of anyons anymore.

\begin{figure}[t]	
 	\centering
 	\includegraphics[trim=5 0 0 0,clip,width=\linewidth]{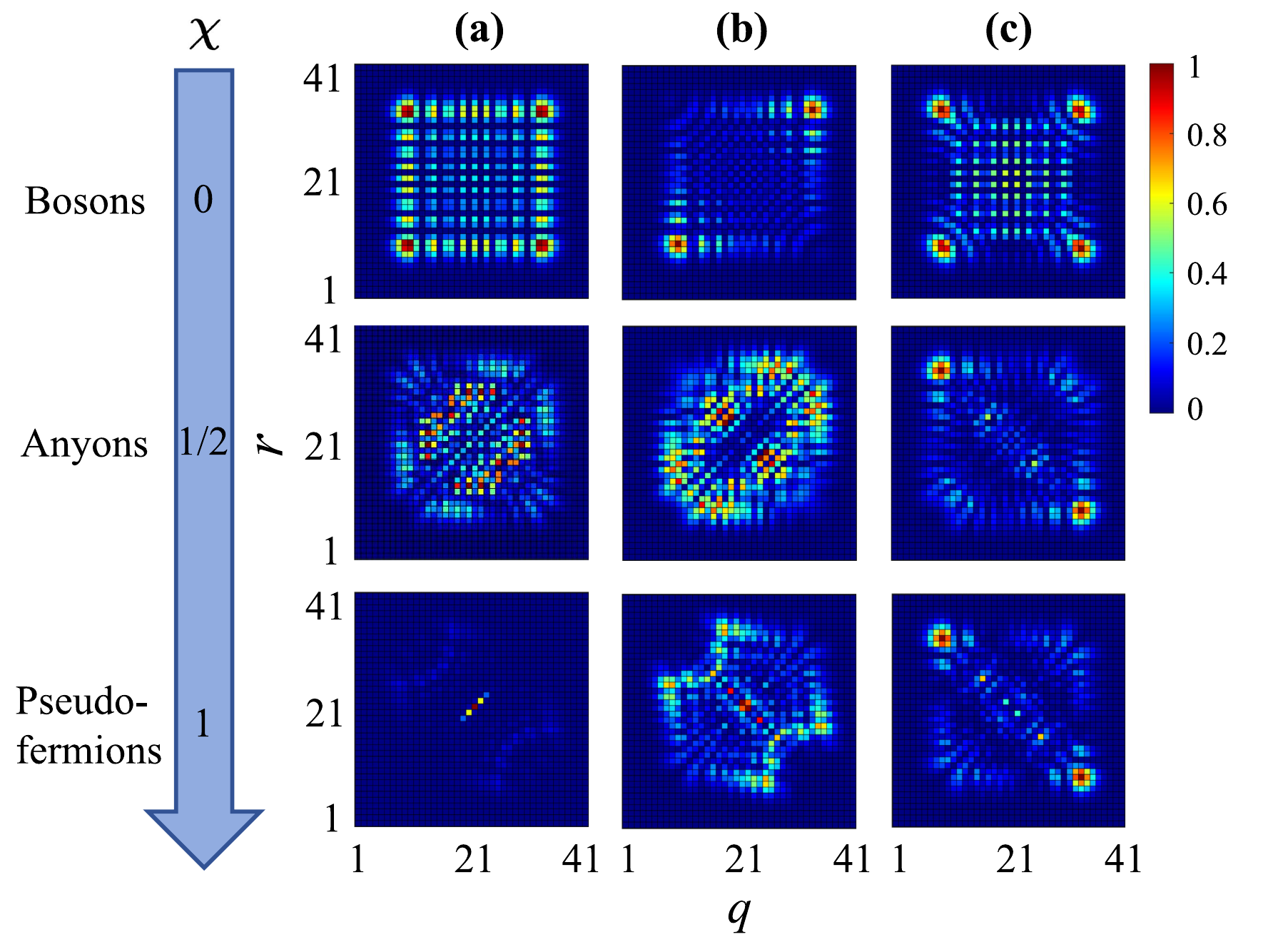}
 	\caption{(Color online) Two-body correlation functions of two anyons in position space on a 1D lattice of $L=41$ sites for external force $F/J = 0.3$ and statistical parameters $\chi = 0, 1/2$ and $1$ from top to bottom. The results calculated in the conventional and orthogonal spaces are exactly the same and we show the correlation at $t=T_B/2$ for three initial states $\left|\phi_{1}\right\rangle$ (a), $\left|\phi_{2}\right\rangle$ (b), and $\left|\phi_{3}\right\rangle$ (c), respectively.} 
  \label{correlation function}
 \end{figure}

Below we continue to consider the correlation function between two anyons with different statistical parameters from three typical initial states in the context of a one-dimensional lattice as shown in Fig. \ref{correlation function}. Anyons with fixed parameter $\chi$ exhibit different correlation behaviors for various initial states, while particles starting from the same initial state may behave distinctly for different $\chi$~\cite{Wang2014}. In the study of BOs, two anyons are initially stacked onto the same lattice site, and only the bosonic and pseudo-fermionic limits are considered in the experiments with electric circuits \cite{Zhang2017}. Here we show that the BOs in the presence of external force depend crucially on both the statistical parameter and the initial states. The correlation functions at a specific time of half BOs period $T_B$ are rescaled by the maximum value such that $\Gamma_{qr}(T_B/2)/\Gamma^{max}_{qr}(T_B/2)$ are presented in Fig. \ref{correlation function} for three initial states introduced above with $F/J=0.3$.  Again we find a perfect coincidence between the conventional and orthogonal schemes for intentionally prepared initial state $\left|\phi_{1,2,3}\right\rangle$ with any statistical parameter $\chi$. Specifically, the case of bosonic limit has been studied in the waveguide lattice system~\cite{Bromberg2009,Alberto2010,Meinecke2013} where the quantum correlations of photon pairs lead to nontrivial quantum interference and strongly depend on the input states. We choose to illustrate the correlation at the time of half the oscillation period $T_B/2$, when the density peaks are far enough away. When two bosons are initially stacked onto the same lattice site, i.e., the $\left|\phi_{1}\right\rangle$ state, the numerical results show no interference occurs between the particles, i.e. the correlation function is just a product of two single-particle probability density. The correlation matrix is characterized by four peaks at the corners of the matrix in the top panel in Fig. \ref{correlation function}(a), which is a result of the ballistic propagation of particles. When the two bosons are prepared in two neighboring sites, i.e., the $\left|\phi_{2}\right\rangle$ state, the correlation map changes considerably as shown in the top panel of Fig. \ref{correlation function}(b). The most obvious feature is the vanishing of the two off-diagonal lobes: the bosons tend to bunch to the same lobe. This can be explained as a generalized Hong-Ou-Mandel interference: two paths lead to a coincidence measurement between site $q$ and site $r$. This interference phenomenon is also known as Hanbury Brown Twiss (HBT) interference\cite{Schellekens2005,Jeltes2007}. The four lobes are recovered when the bosons are initiated in the $\left|\phi_{3}\right\rangle$ state, i.e., with one empty site between the occupied sites. This state contains strong non-classical features with very clear differences from that for the $\left|\phi_{1}\right\rangle$ state as shown in the top panel of Fig. \ref{correlation function}(c). The boson pair exhibits bunching behavior but with a different symmetry: if one particle is detected in between the lobes, the probability of detecting the second one in a lobe vanishes, even though a single particle is most likely to reach the lobes. Similarly, if one boson is detected in a lobe, it is certain that the other one is also in a lobe. We note that the quantum interference emerges since the two bosons are indistinguishable.

In the pseudo-fermionic limit, we observe that starting from the doubly occupied state $\left|\phi_{1}\right\rangle$ the probability of finding both particles at the initial position takes the maximum value as the particles happen to return to and localize at the central lattice site at time $T_B/2$ with a relatively large probability. This is witnessed by the corresponding probability density shown in the lower left panel of Fig. \ref{Pk}. On the other hand, the dynamics of $\left|\phi_{3}\right\rangle$ state shows typical anti-bunching behavior of fermions, i.e. if one anyon is detected at one side, most likely the second one will appear at the other side of the lattice. From the density distribution of the particles, we can see that at the time $T_B/2$, the anyons move to the locations that are the farthest away from the center of the lattice. This is also the case for anyons in the bosonic limit, however, the correlation is completely different. For $\left|\phi_{2}\right\rangle$ the density of the particles is essentially distributed in the central region of the lattice and the correlation reveals fractional statistics, showing both strong diagonal weights and the onset of fermionization. Note that the moment $T_B/2$ is at the first quarter of a full BOs period $2T_B$ in this case. 

For anyons with statistical parameter $\chi=1/2$, interestingly we find the behavior of anyons in state $\left|\phi_3\right\rangle$ resembles that of pseudo-fermions very closely, showing very clear anti-bunching phenomenon at this specific time in first BOs period. This occurs only once as the probability density enters into a chaotic dynamical process without the BOs hereafter. For the other two initial states $\left|\phi_{1,2}\right\rangle$ with $\chi=1/2$, more complicated correlation functions follow the chaotic evolution of the probability density due to the absence of the BOs. When the anyons complete one full BOs period ($T_B$ for bosons and $2T_B$ for pseudo-fermions), either the probability density or the correlation function returns to that of the initial states, respectively, and then the system starts to evolve for another period, which does not happen for anyons for arbitrary intermediate statistical parameters $0<\chi<1$.

\section{Conclusions} 	
In conclusion, we have investigated the stationary and dynamical properties of a pair of non-interacting anyons hopping on a 1D lattice subject to an external force $F$ by expanding the wave function of two-anyon state in the conventional and orthogonal Hilbert spaces, respectively. The eigen equations as well as the time evolution equations in terms of the probability amplitudes are obtained in these two schemes and the coefficients $c_{lk}$ and $c_{kl}$ in the conventional scheme are related by a phase factor due to the anyonic commutation relations. 

The orthogonality of the basis vectors is shown to be very crucial by comparing the stationary energy spectrum and the time-dependent dynamical behavior in the two schemes. The system of two anyons in a 1D lattice can be mapped to a single particle motion in two dimensions and we find the mapping result of the conventional scheme is the entire 2D square lattice region, while that of the orthogonal scheme is only the folded triangular region of the square lattice. Moreover, the conventional scheme produces $L(L-1)/2$ non-physical redundant eigen-states compared to the orthogonal space, whose coefficients violate the anyonic relation and should be excluded from the two-anyon Hilbert space in the related calculations. Our research on the energy spectrum showed that these redundant non-physical eigenvalues affect the overall structure of the energy spectrum, and the number of these non-physical eigenvalues gradually increases with lattice size to converge to the number of physical eigenvalues. 

Finally, by studying the time evolution of the revival probability, the probability density function, and the two-body correlation in these two schemes, we found that the dynamics of the system depends crucially on the initial states. Three initial state with two anyons on the same site, on the neighboring sites, and on the next-neighboring sites with an empty site in between, are taken as examples to illustrate the necessity of the orthogonal basis vectors. We observed exactly the same dynamical behavior in both schemes for initial states with the phase factor $e^{i\chi \pi}$ included in correct positions. However, the calculation for the non-anyonic initial states in the conventional space shows erroneous results, i.e. BO would not occur at $T_B$ for the initial state with two anyons on the (next-)neighboring sites in the fermionic limit, as the conventional scheme predicted. This can be readily check by the electric circuits simulation by preparing the initial state accordingly. In addition, our calculation shows that the two-anyon correlation function recovers the full interference pattern and the bunching behavior in the bosonic limit, which has been realized in the waveguide lattice system. The dynamics of $\left| \phi_3 \right\rangle$ state shows interesting anti-bunching behavior of fermions in the pseudo-fermionic limit as well as in the middle value $\chi=1/2$. 

{\it Acknowledgements}: We thank Prof. Xiang-Dong Zhang for helpful discussions on the selection rule of the energy levels in order to illustrate the BO in the spectrum. This work is supported by National Science Foundation of China (Grant No. 12074340) and the Science Foundation of Zhejiang Sci-Tech University (Grants No. 20062098-Y, 19062463-Y and 22062344-Y). 
\end{document}